\documentclass[twocolumn]{aa}
\usepackage{subcaption}
\usepackage[varg]{txfonts}
\usepackage{graphicx}
\usepackage{multicol}
\usepackage{amsmath}
\usepackage[colorlinks=true,linkcolor=magenta,citecolor=blue,filecolor=cyan,urlcolor=black,final=true]{hyperref}
\usepackage[normalem]{ulem} 
\usepackage{xcolor} 

\begin{document} 

\makeatletter 

\renewcommand*\aa@pageof{, page \thepage{} of \pageref*{LastPage}}
\makeatother

   \title{Structure and fluctuations of a slow ICME sheath observed at 0.5~au by the Parker Solar Probe}
        \titlerunning{Sheath fluctuations}

   \author{E.~K.~J.~Kilpua \inst{1}
        \and
       S.~W.~Good \inst{1}
        \and
       M.~Ala-Lahti \inst{1,2}
        \and
    A.~Osmane \inst{1}    
     \and
      S.~Pal \inst{1}    
     \and
      J.~E.~Soljento \inst{1} 
      \and
      L.L.~Zhao \inst{3}
      \and
    S.~Bale \inst{4}
      }

   \institute{Department of Physics, University of Helsinki, P.O. Box 64, FI-00014 Helsinki, Finland \\
              \email{emilia.kilpua@helsinki.fi}
          \and
          Department of Climate and Space Sciences and Engineering, University of Michigan, Ann Arbor, MI 48109-2143, USA
         \and
          Department of Space Science, The University of Alabama in Huntsville, Huntsville, AL 35899, USA
         \and
          Space Sciences Laboratory, University California, Berkeley, CA 94720, USA
           }

   \date{Received 9 October 2021 / Accepted 21 February 2022}
   
%%%%%%%%%%%%%%%%%%%%%%%%%%%%%%%%%%%%%%%%
%%% ABSTRACT
%%%%%%%%%%%%%%%%%%%%%%%%%%%%%%%%%%%%%%%%
 
  \abstract
  % context 
    {Sheath regions ahead of interplanetary coronal mass ejections (ICMEs) are compressed and turbulent global heliospheric structures. Their global and fine-scale structure are outstanding research problems, and only a few studies have been conducted on this topic closer to the Sun than 1 au. Comprehensive knowledge of the sheath structure and embedded fluctuations and of their evolution in interplanetary space is important for understanding their geoeffectiveness, their role in accelerating charged particles to high energies, the interaction of ICMEs with the ambient wind, and the transport of energy between boundaries.}
  % aims
    {Our key aims are to investigate in detail the overall structure, as well as nature (stochastic, chaotic, or periodic) and origin, of magnetic fluctuations within a sheath ahead of a slow ICME in the inner heliosphere.}
  % methods
    {We used magnetic field and plasma observations from the Parker Solar Probe (PSP) during a sheath region observed at $\sim 0.5$ au on March 15, 2019, ahead of a slow and slowly expanding streamer blow-out CME bracketed between a slower and faster stream. To examine the magnetohydrodynamic-scale turbulent properties, we present an analysis of the fluctuation amplitudes, magnetic compressibility of fluctuations, partial variance of increments (PVI), normalised cross helicity ($\sigma_c$), and normalised residual energy ($\sigma_r$). We also conducted a Jensen-Shannon permutation entropy and complexity analysis.}
  % results
    {The investigated sheath consisted of slower and faster flows that were separated by a brief ($\sim 15$ min) change in the magnetic sector bounded by current sheet crossings and a velocity shear zone. The fluctuation amplitudes and frequency of high PVI values were larger and higher throughout the sheath than in the upstream wind and had dominantly negative $\sigma_r$ and strongly positive $\sigma_c$.  The velocity shear region marked a strong increase in temperature and specific entropy, and the following faster flow had large local patches of positive $\sigma_r$ as well as larger fluctuation amplitudes and higher PVI values, in particular at smaller timescales.  Fluctuations in the preceding wind and in the sheath were found to be stochastic. However, sheath fluctuations showed lower entropy and higher complexity, with entropy showing a reducing and complexity an increasing trend with increasing time lag.}
  % conclusions
    {The two-part sheath structure was likely a result of a warp in the heliospheric current sheet (HCS) that was swept up and compressed into the sheath. The driving ejecta accelerated and heated the wind at the back of the sheath, which then interacted with the slower wind ahead of the HCS warp. This also caused some distinct differences in fluctuation properties across the sheath. Sheaths of slow ICMEs originating as streamer blow-outs can thus have complex structure where fluctuation properties are not just downstream shock properties, but are generated within the sheath. At short timescales, fluctuations feature fully developed and imbalanced MHD turbulence, while at longer scales, fluctuations are increasingly dominated by intermittent coherent and ordered structures.}


\maketitle

%%%%%%%%%%%%%%%%%%%%%%%%%%%%%%%%%%%%%%%%
%%% INTRODUCTION
%%%%%%%%%%%%%%%%%%%%%%%%%%%%%%%%%%%%%%%%

\section{Introduction}

{Sheath regions \citep[e.g.][]{Kilpua2017} ahead of interplanetary coronal mass ejections (ICMEs) are large-scale heliospheric structures that are of considerable interest as drivers of geomagnetic storms \citep[][]{Tsurutani1988,huttunen2002,huttunen2004,Kilpua2017SSR} and as drivers of dramatic variations in the radiation belts surrounding Earth \citep[e.g.][]{Hietala2014,Kilpua2015,Alves2016,Lugaz2016,Kalliokoski2020}. They may also have a significant role in the energisation and transport of charged particles in the corona and interplanetary space \citep[e.g.][]{Manchester2005,Perri2021,Kilpua2021b}. A key feature of sheaths important for space weather considerations is their turbulent, compressed nature \citep{Kilpua2019}. In particular, sheaths have enhanced magnetic fluctuation amplitudes and a greater prevalence of coherent intermittent structures when compared to their surroundings \citep{Kilpua2013,Moissard2019,Good2020a,Kilpua2020,Kilpua2021}.}

{However, the structure and precise nature and origin of magnetic fluctuations in sheaths is still an open and relatively little studied question. A number of studies have investigated the downstream region of interplanetary shocks \citep[e.g.][]{Kajdic2012,Pitna2016,Borovsky2020,Zhao2019a,Zhao2019b,Zank2021}, but considerably fewer studies have probed fluctuations in the ICME sheath holistically. Sheaths pile up gradually as the structured solar wind ahead is first processed by the ICME leading shock and are then compressed into the sheath. Therefore, pre-existing structures in the preceding wind, such as discontinuities, small-scale flux ropes, magnetic reconnection exhausts, and magnetic holes may be swept into the sheaths. Fluctuations can also be further processed and generated in the sheath. Despite propagating super-Alfvénically through the solar wind, ICMEs typically expand strongly in interplanetary space, forcing plasma to pile up at its leading edge rather than flowing around \citep{Siscoe2008}. The interplanetary magnetic field (IMF) also is draped about the ICME \citep[e.g.][]{Comas1988}, generating strong out-of-ecliptic fields in the sheath. The shock and ICME expansion (and associated field line draping) provide free energy for the generation of different plasma waves, for example, Alfvén ion cyclotron and mirror mode waves \citep[][]{Alalahti2018,Alalahti2019}.} 

{It is also noteworthy that different parts of the sheath present solar wind that was transmitted through the shock and processed by it at different times. Previous studies exploring turbulent properties in different subregions of the sheath have also revealed that fluctuation properties vary significantly from the shock to the ICME leading edge \citep[][]{Kilpua2020,Kilpua2021}. However, the lack of a $1/f$ spectrum in the magnetohydrodynamical (MHD) inertial range  \citep{Kilpua2021}, in contrast to planetary magnetosheaths \citep[e.g.][]{Huang2016,Hadid2015}, suggests that ICME sheaths do not strongly reset the turbulence, but rather amplify it. It is therefore likely that the key differences predominantly arise from processes internal to the sheath.}

{In this paper, we analyse a sheath region that was related to an ICME encountered by the Parker Solar Probe \citep[PSP;][]{Fox2016} at a heliospheric distance of 0.547 au from the Sun. This event has been analysed by \cite{Lario2020}, who focused on the ICME flux rope structure and energetic particle acceleration. The ICME was associated with a slow streamer blow-out CME \citep[e.g.][]{Sheeley1982,Vourlidas2018} that was subsequently accelerated by a trailing fast stream, up to a speed high enough to produce a pair of relatively weak and slow leading shocks. Streamer blow-out CMEs originate as disruptions of the the coronal streamer belt and are characterised by slow evolution, slow speeds, and large angular widths. While fast ICMEs have generally stronger shocks and more prominent sheaths, slower ICMEs can also have significant sheaths that are important for space weather \citep{Kilpua2019} and particle energisation \citep{Kilpua2021}.}

{The PSP high-cadence magnetic field measurements and the availability of plasma data allow a detailed investigation of a streamer blow-out ICME sheath closer to the Sun than has been performed in most previous studies. The previous work by \cite{Good2020a} analysed a slow ICME sheath at MESSENGER at 0.47 au using magnetic field measurements, featuring distinct fluctuation properties in comparison to the solar wind ahead. In addition, we here explore how the fluctuation properties vary within the sheath using a sliding-window average.} 

{The paper is organised as follows: In Section~\ref{sec:analysis} we describe the data and give an overview of the event. In Section~\ref{sec:results} we describe the results of the fluctuation analysis. Section~\ref{sec:discussion} discusses the results, and conclusions are given in Section~\ref{sec:conclusion}.}

%%%%%%%%%%%%%%%%%%%%%%%%%%%%%%%%%%%%%%%%
%%% OBSERVATIONS & DATA ANALYSIS
%%%%%%%%%%%%%%%%%%%%%%%%%%%%%%%%%%%%%%%%

\section{Data and event overview} \label{sec:analysis}

\subsection{Data}

We used observations from the PSP. Magnetic field measurements come from the FIELDS instrument suite \citep{Bale2016} at a resolution of 0.128~s, and plasma data come from the Solar Wind Electrons Alphas and Protons (SWEAP) instrument suite \citep{Kasper2016} at a resolution of 27.96~s. The data were achieved as level 2 data through the public automated multi-dataset analysis (AMDA; \url{http://amda.irap.omp.eu) database. }

\subsection{Event overview}

{The event studied here is the sheath region observed on March 15, 2019. The overall properties of this event have been analysed in detail by \cite{Lario2020}, who focused on the shock and flux rope properties, as well as on energetic particles. The authors noted two closely spaced ($\sim 4$ min apart) interplanetary shocks that likely resulted from the interaction of the ICME with the following faster stream. The first shock occurred on March 15 at 08:56:01 UT and the second at 09:00:07 UT. The authors applied Rankine–Hugoniot conservation equations to plasma and magnetic field data to derive key shock parameters. For the first shock, they obtained an angle between the shock normal and the upstream field $\theta_{Bn}=46 \pm 2^{\circ}$, a shock speed  $V_{sh} = 370 \pm 135$ km/s, and a fast magnetosonic Mach number $M_{ms} = 2.0 \pm 2.4$. For the second shock, the corresponding values are $\theta_{Bn}=60 \pm 2^{\circ}$,  $V_{sh} = 376 \pm 34$ km/s, and $M_{ms} = 2.1 \pm 0.3$. Both shocks were thus relatively weak and slow and had quite similar properties. The second shock was also slightly faster and stronger and had higher density and field compression ratios \citep{Lario2020}.}

%%%%% figure 1 %%%%%
\begin{figure}[ht]
\centering
\includegraphics[width=0.99\linewidth]{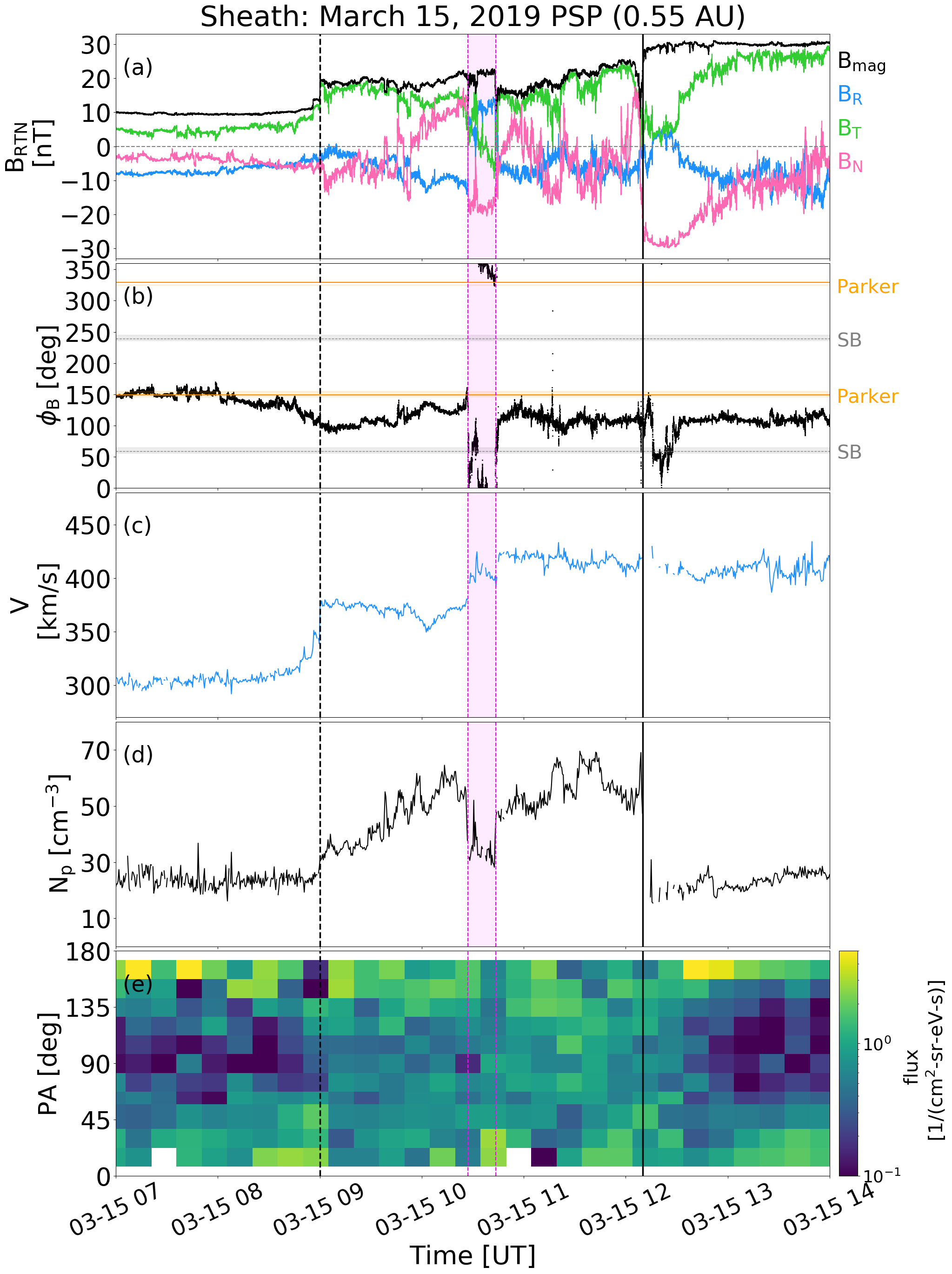}
\caption{Solar wind magnetic field, plasma, and suprathermal electron data as measured by the PSP during the ICME-driven sheath region on March 15, 2019. From top to bottom: a) Magnetic field magnitude (black) and three magnetic field components in RTN (blue: $B_R$, green: $B_T$, pink: $B_N$). b) Azimuth RTN angle of the magnetic field. c) Solar wind velocity R-component. d) Solar wind density. e) Normalised (to the mean of each time) suprathermal electron pitch angle distributions for the energy range 352.9 - 438.8 eV. The dashed vertical line shows the shock, and the solid vertical line shows the ICME leading edge. Two vertical dashed pink lines and the pink shaded area in between mark the region between the two reconnection exhausts (see text for details). The horizontal orange lines in panel 1b) mark the Parker spiral directions calculated for the average solar wind speed during the investigated interval, and the shaded region indicates the range obtained using the minimum and maximum speeds. The grey horizontal lines in panel 1b) show the SBs between towards and away sectors.}
\label{fig:overview}
\end{figure}
%%%%%%%%%%%%%%%%%%%%

{The overview of the sheath is shown in Figure~\ref{fig:overview}. The dashed vertical line shows the second shock on March 15, 2019, at 09:00:07 UT, and the solid vertical line shows the leading edge of the ICME detected at 12:10 UT. The sheath lasted 3.16 hr (190 min). Visual inspection of the data shows that the sheath features higher density when compared to the surroundings.}

{The solid purple lines in panel \ref{fig:overview}b show the RTN longitudes ($\phi_B$) that correspond to the inward and outward Parker spiral angles of $149^{\circ}$ and $329^{\circ}$ at 0.55 au (acute spiral angle $31^{\circ}$), calculated using the average solar wind speed of 373 km/s during the interval displayed. The dashed grey lines show the boundaries between the toward ($59^{\circ} < \phi_B < 239^{\circ}$) and away magnetic sectors. The shaded areas show the variations in Parker spiral angle and sector boundaries (SB) when the spiral angle is calculated using the minimum (305 km/s) and maximum (425 km/s) speeds during the shown interval. An SB crossing from the away to the toward sector occurred on March 14, about half a day before the leading shocks of the flux rope (data not shown here; see e.g. Figure 1 in \cite{Lario2020}). Before the shock, the IMF was thus in the toward sector and aligned with the Parker spiral, as shown in Figure~\ref{fig:overview}b. In the sheath, the IMF stayed mostly in the toward sector, but deviated from the spiral direction. The bottom panel of suprathermal electrons indicates that the strongest heat flux in the preceding wind and during the sheath was mostly at pitch angles (PA) $\sim 180^{\circ}$, that is, flowing anti-parallel to the magnetic field.}

{The most striking feature in the sheath is the sharp change in $B_N$ marked by the first dashed vertical magenta line. The zoom-in to this feature (see Figure~\ref{fig:Exhaust1} in the appendix) reveals that it is likely a reconnection exhaust \citep[e.g.][]{Gosling2006}. Between 10:27:04 UT and 10:27:29 UT, a step-like change in the magnetic field direction and a simultaneous strong dip in the magnetic field magnitude is observed. However, the plasma data have a gap around this time, so that the interpretation cannot be further confirmed. The second magenta line  (see Figure~\ref{fig:Exhaust2} in the appendix) shows another possible reconnection exhaust between 10:43:11 UT - 10:44:53 UT with the clearest change in $B_T$, but the structure is not as sharp as the first. The IMF between the exhausts is in the away sector, and the interval displays a steady and strongly negative $B_N$, a reduced density, and marks a $\sim 50$ km/s increase in solar wind speed from about 375 km/s to 425 km/s. We discuss this region in more detail in Section~\ref{sec:sigr_sigc}.}

{The leading edge of the flux rope is clear (solid vertical black line). The transition from the sheath to the flux rope features an abrupt change in the magnetic field direction, the start of the smooth rotation in the magnetic field direction, and a clear drop in plasma density. This transition from the sheath to the flux rope is not instantaneous, but occurs within an approximately 5-min boundary layer. Such layers are typical at the boundaries of interplanetary magnetic flux ropes, likely generated by interactions in interplanetary space, or they may be remnants of the CME release process at the Sun \citep[e.g.][]{Wei2003}.}

%%%%%%%%%%%%%%%%%%%%%%%%%%%%%%%%%%%%%%%%
%%% RESULTS
%%%%%%%%%%%%%%%%%%%%%%%%%%%%%%%%%%%%%%%%

\section{Results}
\label{sec:results}

\subsection{Magnetic fluctuations}
\label{sec:mag_fluc}

\begin{figure}[ht]
\centering
\includegraphics[width=0.99\linewidth]{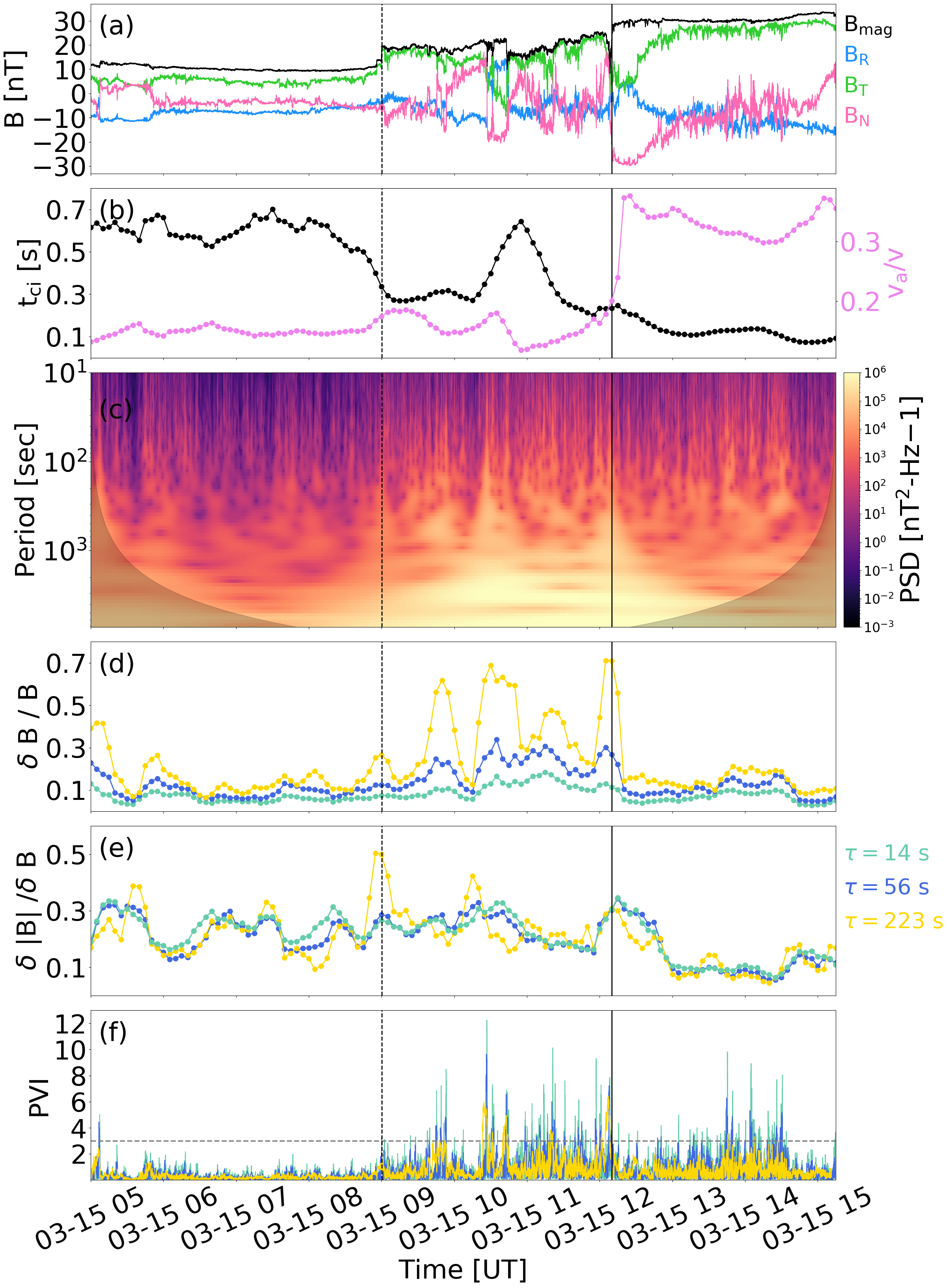}
\caption{Magnetic field fluctuations during the sheath on March 15, 2019. From top to bottom, we show a) the magnetic field magnitude (black) and three magnetic field components in RTN (blue: $B_R$, green: $B_T$, pink: $B_N$), b) 
the ion cyclotron timescale ($t_{ci}$) and the ratio of the Alfv\'{e}n speed to solar wind speed ($v_A / v$), c) the wavelet PSD of magnetic field fluctuations in the periods of 1 s to 2 hrs ($\sim 0.1$ mHz - 1 Hz frequency range), d) the normalised magnetic field fluctuations ($\delta B / B$), e) the fluctuation compressibility ($\delta |B| / \delta B$), and f)  the PVI. The properties at three different MHD timescales (14 s, 56 s, and 223 s) are shown in panels d)-f). In panels b), d), and e), values are calculated with a 20-min window sliding in 5-min steps through the investigated interval. The horizontal line in panel f) marks the value PVI$=3$. The vertical dashed line marks the shock, and the solid line shows the ICME leading edge, similar to Figure~\ref{fig:overview}.}
\label{fig:wavelet}
\end{figure}

{Figure~\ref{fig:wavelet} explores magnetic fluctuations in the sheath. The top panel of Figure~\ref{fig:wavelet} repeats the magnetic field observations from Figure~\ref{fig:overview}, and the second panel examines the expected physical range of fluctuations and the validity of the Taylor hypothesis. The data points are calculated as 20-min sliding means in 5-min steps. The black dots show the spacecraft-frame timescale, $t_{ci}$, corresponding to the proton gyroradius length scale,

\begin{equation}
t_{ci}=\frac{\langle v_{th} \rangle}{\langle v \rangle} \frac{2 \pi m_i}{e \langle B \rangle}
\end{equation}

where $\langle v \rangle$, $\langle v_{th} \rangle$, and $\langle B \rangle$ are the averages of the solar wind speed, thermal speed, and magnetic field magnitude, respectively. The timescale $t_{ci}$ is associated with the spectral break between the MHD and kinetic ranges with increasing plasma-$\beta$ \citep{Chen2014}. Figure~\ref{fig:overview}b shows that $t_{ci}$ values for the studied interval are all below 0.8~s; we here study fluctuations at scales of 
$\sim 10$~s and above, which are thus expected to be in the MHD range. The purple dots in the same panel show the ratio of the Alfv{\'e}n speed $v_A$ to the solar wind speed $v$. This ratio can be considered as a validity test for the Taylor hypothesis, which states that the path taken by a spacecraft through the solar wind can be considered as an instantaneous spatial cut if the timescales of magnetic field fluctuations are sufficiently shorter than the timescale of the solar wind flow \citep{Taylor1938,Matthaeus1982}. For the Taylor hypothesis to be valid at MHD scales, $v_A /v \lesssim 1$ \citep{howes2014}. This criterion is clearly met during the investigated interval.}

%%%%% Figure 2 %%%%%
%%%%%%%%%%%%%%%%%%%%

{Panel c) in Figure~\ref{fig:wavelet} shows the wavelet power spectral density (PSD) of magnetic field fluctuations with periods from 10 s to 2 hrs ($\sim 0.01$ mHz - 1 Hz). The PSD shows that the fluctuation amplitudes across this range are enhanced in the sheath when compared to the surrounding (or preceding) wind, particularly when compared to the preceding wind. The leading part of the flux rope has some larger amplitude fluctuations, most distinct in $B_N$, but we do not discuss them further here.}

{The next two panels (d and e) show the magnetic field fluctuations normalised to the mean field ($\delta B / B$) and fluctuation compressibility ($\delta |B| / \delta B$). The magnetic field fluctuations are defined here as $\delta \mathbf{B} = \mathbf{B}(t) - \mathbf{B}(t+\tau)$, where $\tau$ is the fluctuation timescale, that is, the time lag between two sample points, and the amplitude of the fluctuation is $\delta B = |\delta \mathbf{B}|$.  We calculated $\delta B / B$ and $\delta |B| / \delta B$ for three timescales (14, 56, and 223 s), and similarly as $t_{ci}$ and $v_A/v$ in panel b),  as the mean over a 20-min window sliding in 5-min steps.} 

{The levels of normalised magnetic field fluctuations are higher in the sheath than in the surroundings for all timescales when compared to the preceding wind. As the timescale $\tau$ increases,  $\delta B / B$ becomes higher than in the upstream wind. As the spectral slope can be calculated as the gradient of $\delta B / B$, this implies that the MHD range spectral slopes are steeper in the sheath than in the preceding  wind.  Overall, $\delta B / B$ values show large variations throughout the sheath. For $\tau = 223$ s, the jumps in $\delta B / B$ are the most prominent, and the highest $\delta B / B$ values coincide with the sharp field changes within the sheath, shock, and the ICME leading. For $\tau = 0.14$ s and 56 s, there is not much change across the shock, and the $\delta B / B$ values tend to be higher in the trailing part of the sheath. The magnetic compressibility of fluctuations in turn did not increase from the upstream wind to the sheath. The $\delta |B| / \delta B$ values stayed at about the same level in the front part of the sheath and showed a declining trend towards the ICME leading edge in the trailing part of the sheath. When entering the flux rope, $\delta |B| / \delta B$ clearly dropped, consistent with previous studies \citep[][]{Moissard2019}.}  

{The partial variance of increments (PVI) technique allows locating sharp field discontinuities in time series that contribute to the “fatter” non-Gaussian tails of probability density functions (PDFs) of fluctuations \citep[e.g.][]{Greco2008,Zhou2019,Zhao2020}. These structures can arise spontaneously from turbulence or be inherent solar wind structures (e.g. current sheets or flux tube boundaries).  The PVI parameter is defined as}
\begin{equation}
PVI = \frac{|\delta \mathbf{B}|}{\sqrt{\langle|\delta \mathbf{B}|^2\rangle}}, 
\label{eq:pvi}
\end{equation}
{where  $\delta \mathbf{B}$ is the field increment, and the average in the denominator is taken over the interval shown in Figure~\ref{fig:wavelet}. Values of PVI $> 3$  are typically considered to indicate the presence of intermittent coherent structures. The PVI time series in the bottom panel of Figure \ref{fig:wavelet}f show that for all timescales, the PVI values are clearly lower in the solar wind preceding the shock than in the sheath. Large PVI peaks (PVI $> 3$) in the sheath mostly occur for $\tau = 14$ and 56 s and are most frequent in the middle and latter half of the sheath.}

\begin{figure}
\captionsetup[subfigure]{labelformat=empty}
\centering
   \begin{subfigure}[b]{0.99\textwidth}
   \includegraphics[width=0.50\textwidth]{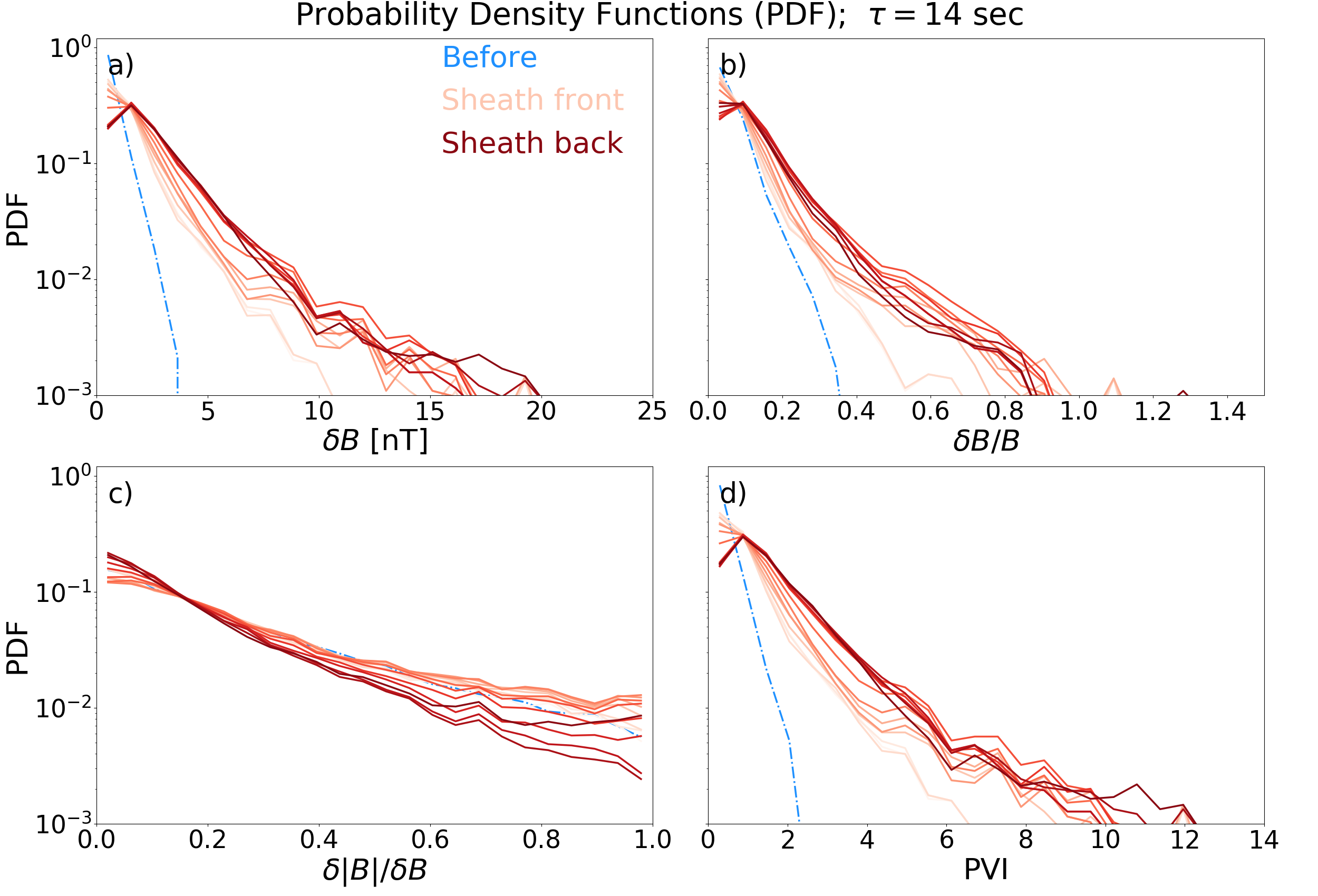}
   \caption{}
   \label{fig:Ng1} 
\end{subfigure}
   \\
\begin{subfigure}[b]{0.99\textwidth}
   \includegraphics[width=0.50\textwidth]{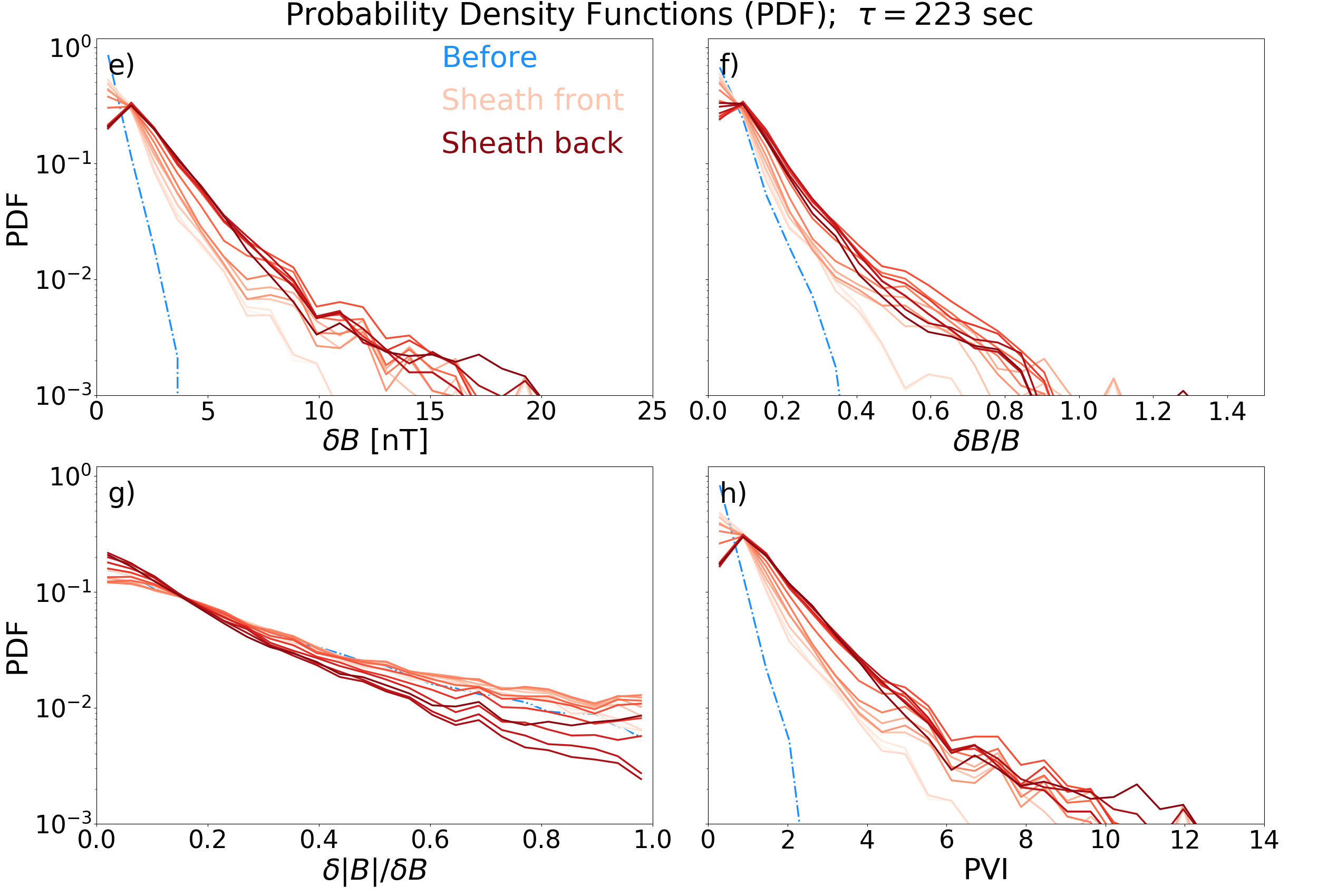}
   \caption{}
   \label{fig:Ng2}
\end{subfigure}
\caption{Probability density functions in panels a) and e) of the fluctuation amplitude ($\delta B$), in panels b) and f) for the normalised fluctuation amplitude ($\delta B/B$), in panels c) and g) for the magnetic compressibility of fluctuations ($\delta |B|/\delta B$), and in panels d) and h) for the PVI for the timescale (top) 14 s and (bottom) 223 s. The blue histograms show the results for the 1-hr interval of the solar wind before the shock, and the light orange to dark red curves for 1-hr intervals in the sheath. The shade of red darkens from the shock to the ICME leading edge.}
\label{fig:distributions}
\end{figure}

{Figure~\ref{fig:distributions} shows PDFs of $\delta B$, $\delta B / B$, $\delta |B| / \delta B$ and PVI for  $\tau = 14$ s and  $\tau = 223$ s in the preceding solar wind and in the sheath. In the sheath, PDFs were calculated over a 1-hr interval sliding in 15-min steps throughout the sheath. First, the figure shows that the  $\delta B$, $\delta B / B$, and PVI distributions have considerably fatter tails in the sheath than in the solar wind ahead. For $\tau = 14$ s,  the PDF tails are thinnest at the sheath front, but otherwise, no clear trend is seen. For $\tau = 223$ s, the highest values occur in the trailing part of the sheath.  The PDFs of magnetic compressibility $\delta |B| / \delta B$ are in turn similar between the preceding wind and sheath. For both $\tau = 14$ s and  $\tau = 223$ s, the compressibility values are generally lower in the trailing part of the sheath, as noted previously.} 

%%%%% figure 3 %%%%%
%%%%%%%%%%%%%%%%%%%%

\subsection{Residual energy and cross-helicity}
\label{sec:sigr_sigc}

\begin{figure}[ht]
\centering
\includegraphics[width=0.99\linewidth]{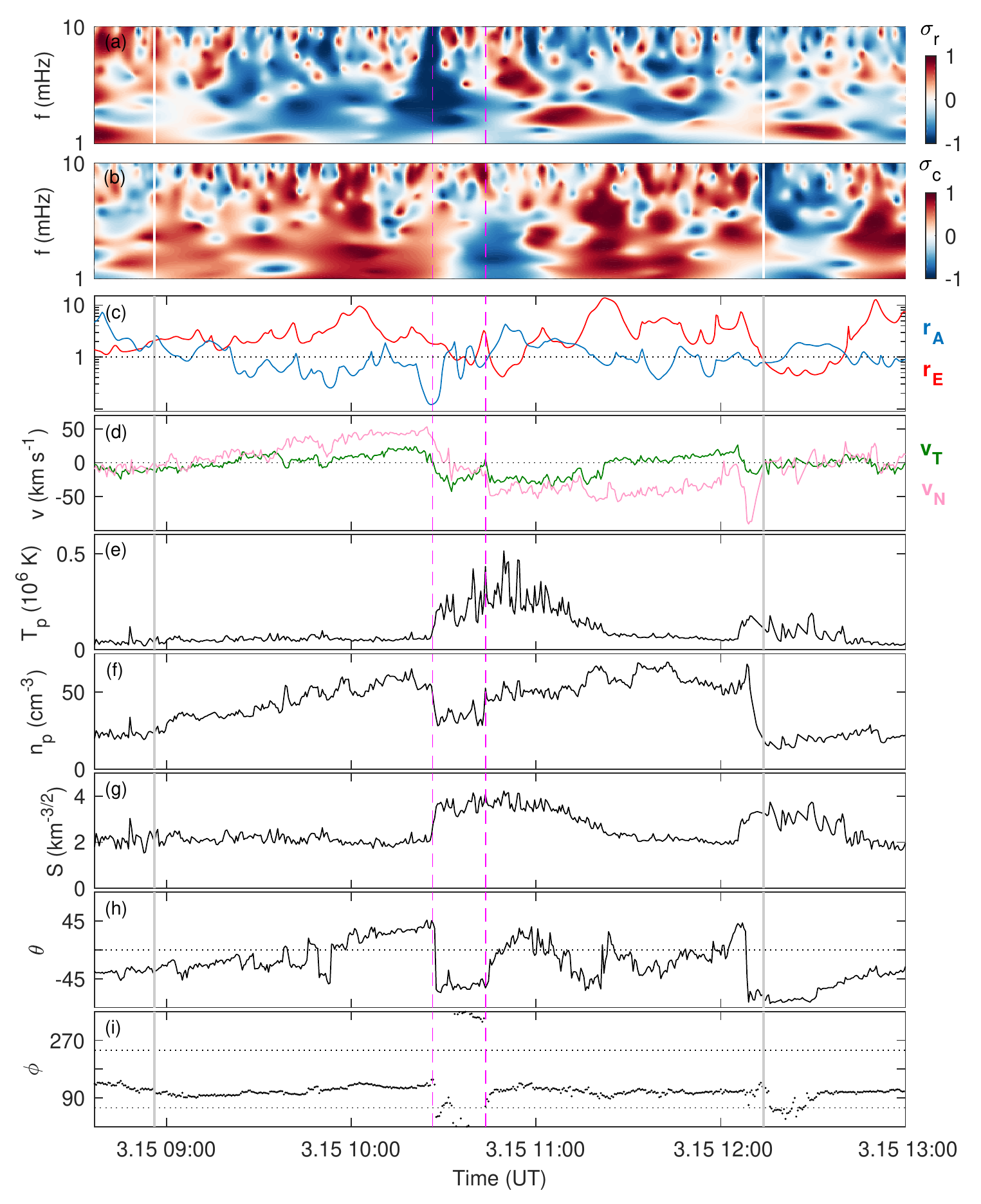}
\caption{Various parameters relating to the sheath fluctuations. a) Normalised residual energy; b) normalised cross helicity; c) Alfv\'en and Elsässer ratios; d) $T$ and $N$ velocity components; e) proton temperature; f) proton density; g) specific entropy; h) RTN magnetic field latitude angle; and i) RTN magnetic field longitude angle, where dotted horizontal lines indicate the nominal SBs. Vertical grey lines denote the sheath boundaries, and two pink lines show the reconnection exhausts.}
\label{fig:elsasser}
\end{figure}

{The normalised residual energy, $\sigma_r$, and normalised cross-helicity, $\sigma_c$, of MHD-scale fluctuations in the frequency range $10^{-3} - 10^{-2}$~Hz (100 s to 16.7 min) were also investigated; see panels a) - b) in Figure~\ref{fig:elsasser}.  The analysis of these quantities can give particular insight into the interaction of the ICME with the surrounding solar wind and into the origin of the fluctuations. The normalised residual energy is a widely used measure to determine the degree to which fluctuations are Alfv\'{e}nic. Values of $\sigma_r > 0$ indicate excess energy in velocity fluctuations, and $\sigma_r < 0$ indicates excess energy in magnetic field fluctuations. For Alfv\'{e}n waves, it is predicted that $\sigma_r \approx 0$. The normalised cross-helicity indicates whether  there is more power in Alfv\'{e}nic wave packets propagating parallel ($\sigma_c < 0$) or anti-parallel ($\sigma_c > 0$) to the mean magnetic field, or if they have approximately equal power ($\sigma_c \approx 0$). When $\sigma_c \approx 0$, the turbulence is said to be balanced, or it is an indication that the fluctuations are not Alfv\'{e}nic.} 

{Here, the normalised residual energy and cross-helicity are determined from the trace Morlet wavelet power spectra $E_v$, $E_b$, and $E_{\pm}$ \citep[e.g.][]{Chen2013,Good2020b,Zhao2020b,zhao2021}, expressed as}
\begin{equation}
\sigma_r = \frac{E_v - E_b}{E_v + E_b}\,
\end{equation}
and
\begin{equation}
\sigma_c = \frac{E_+ - E_-}{E_+ + E_-}\,
,\end{equation}
{where ${\boldsymbol{v}}$ is the plasma velocity, ${\boldsymbol{b}}={\boldsymbol{B}}/\sqrt{{\mu }_{0}\rho }$ is the magnetic field given in velocity units, with $\rho$ denoting the particle density, and ${{\boldsymbol{z}}}^{\pm }\,={\boldsymbol{v}}\pm {\boldsymbol{b}}$ are the Elsässer variables, which define the anti-parallel (+) and parallel (-) propagating wave packets. Magnetic field and plasma data were resampled to a resolution of 30~s for this analysis. The fluctuations in this analysis are assumed to be incompressible. The spectrograms of $\sigma_r$ and $\sigma_c$ across the sheath interval are shown in Figure~\ref{fig:elsasser}. We also show in Figure~\ref{fig:elsasser} timeseries of the Alfv\'{e}n ratio $r_A = E_v/E_b$ and Elsässer ratio $r_E = E_+ / E_-$, averaged over $10^{-3} - 10^{-2}$~Hz. Similar in nature to $\sigma_r$ and $\sigma_c$, these ratios give more of an absolute rather than normalised measure of balance or imbalance.}

{The sheath mostly consisted of fluctuations with negative $\sigma_r$ ($\equiv r_A < 1$), indicating the dominance of the magnetic fluctuations, with some significant localised patches of positive $\sigma_r$ ($\equiv r_A > 1$) just downstream of the shock and after the second magenta line, that is, in the trailing part of the sheath. The sharp magnetic field discontinuity indicated by the first magenta line is the cause of the strongly negative $\sigma_r$ (with minimum $r_A\sim 0.12$) at this time.  The regions in which $\sigma_r \approx 0$ are found are close to the shock and ICME leading edge.}

{The cross-helicity $\sigma_c$ was mostly positive ($\equiv r_E > 1$) in the sheath, indicating that more power was present in wave packets propagating anti-parallel to the magnetic field. Since the magnetic field was almost entirely in the toward sector (the magnetic field latitude angle $\phi$ fell between the dotted horizontal lines in Figure~\ref{fig:elsasser}i, which give the SBs for a Parker spiral angle of $31^{\circ}$), this dominant anti-parallel flux was thus anti-sunward. The field briefly veered into the nominal away sector between the magenta lines, partly coinciding with a reversal of the cross-helicity sign.}

{We note that the large-scale velocity gradient and a considerable increase in  solar wind temperature present in the middle of the sheath associated with HCS crossings coincide with a region where $\sigma_r$, $\sigma_c$, $r_A$, and $r_E$ deviate from the values typically seen elsewhere in the sheath. This shear zone, as seen in $V_T$ and $V_N$ in Figure~\ref{fig:elsasser}d, preceded an increase in the bulk solar wind speed shown in Figure~\ref{fig:overview}c. This region featured strongly negative $\sigma_r$ and low $r_A$, suggesting a strong dominance of energy in the magnetic fluctuations. It is followed by a period with localised positive $\sigma_r$ patches and high $r_A$, suggesting in turn that energy dominates in the velocity fluctuations. The $\sigma_c$ in turn show generally more $\sim 0$ and $r_E \sim 1$ values than elsewhere in the sheath, indicating waves propagating both parallel and anti-parallel to the background field. 
Figure~\ref{fig:elsasser}d shows that the first HCS crossing marks a region of strongly enhanced temperature. In the front part of the sheath, the temperature is relatively low as it enhances only slightly at the shock. The solar wind density is in turn strongly enhanced in the sheath, as previously mentioned in Section~\ref{sec:mag_fluc}, and it dips between two HCS crossing when the magnetic sector changes (Figure~\ref{fig:elsasser}e). The first HCS crossing also marks a clear increase in specific entropy (Figure~\ref{fig:elsasser}g, defined here as $S = \ln{T_p/N_p^{1/2}}$; \citet[][]{Pagel2004}). Changes in $S$ may be related to localised heating by dissipative processes at current sheets, velocity shears, and shocks \citep[e.g.][]{Borovsky2010}, or to changes in heavy ion charge-state ratios \citep[e.g.][]{Pagel2004}. The velocity shear, the increase in proton temperature and specific entropy, and the drop in proton density are all signatures of a stream interface (SI) separating the slow and fast wind \citep[e.g.][]{Borosvky2008}. Although the HCS and SI are typically separated by at least a few hours, they can also coincide \citep[e.g.][]{Huang2016}. The enhanced density also suggests that the studied sheath  piled up from the material in the heliospheric plasma sheet.  We note that there was also a strong velocity deflection just before the ICME leading edge as seen in $V_T$ and $V_N$.}

\subsection{Complexity - entropy analysis}

{Finally, we investigated the nature of fluctuations in the sheath using the permutation entropy methodology introduced by \cite{Bandt2002} and the Jensen-Shannon complexity analysis proposed by \cite{Rosso2007}. These approaches have been widely used in different contexts. In space plasma physics, they have been applied to analyse geomagnetic indices \citep{Osmane2019} and fluctuations in the solar wind \citep{Weck2015,Weygand2019,Good2020a}}, for instance.

{The permutation entropy is determined by investigating the occurrence of patterns in an evenly sampled time series. A pattern consists of $d$ subsequent data points separated by a timescale $\tau$. The number of data points in a pattern, that is, $d$, is called the embedded dimension, and the factorial of $d$ ($d!$) gives the number of possible permutations. In a time series consisting of $N$ data points, the total number of patterns having embedded dimension $d$ and considering timescale $\tau$ is $N-(d-1) \tau$.}

{Let $P$ be the probability distribution of a set of patterns and $p_i$, where $i = 1,2,...,d!$, the probability of an occurrence for a permutation $i$. The permutation entropy is then determined as Shannon’s information entropy,}

\begin{equation}
 S(P) = -  \sum_{i=1}^{d!} p_j \log{p_j,}
\end{equation}
{and the normalised permutation entropy is}

\begin{equation}
 H(P) = -  S(P)/\log{d!.}
\end{equation}

{To compare permutation entropies that have different dimensions, it is convenient to normalise $S(P)$ with $d-1$. This is called permutation entropy per symbol:}

\begin{equation}
H_n(P) = -  S(P)/(d-1).
\end{equation}

{The Jensen-Shannon complexity is defined as

\begin{equation}
C_J^S = - 2 \frac{S(\frac{P+P_e}{2})-\frac{1}{2} S(P) - \frac{1}{2} S(P_e)}{\frac{d!+1}{d!} \log{(d!+1)} - 2 \log{(2d!)} + \log{d!}} H(P).
\end{equation}

In the above, $P_e$ is the probability distribution that maximises the Shannon entropy. This occurs when all patterns are equally likely to occur, that is, they have a probability $p_i = 1/d!$. }

{The statistical robustness of the complexity-entropy analysis is satisfied when  $N/d! > 10$ and $\sqrt{d!/N-(d-1) \tau} < 0.2$, where $N$ is the number of samples in the time series \citep[e.g.][]{Osmane2019}. In the analysis, we analysed both  3-hr and 1-hr times series. For 1s resolution, the former thus have 10800 samples and the latter have 3600 samples. For the 3-hr time series, we varied $\tau$ from 20 s to 1200 s, and for the embedded dimension, we varied from 3 to 5 s, while for the 1-hr time series, $\tau$ ranged from 20 s to 400 s. We investigated only the embedded dimension 4.  For all these cases, the robustness criteria above are met. We also used $\sqrt{d!/N-(d-1) \tau}$, which is related to the average permutation occupation number \citep[e.g.][]{Weygand2019,Good2020a}, as the estimate of the uncertainty range in the time series. For the 3-hr time series and for $d=3,$ the error is about 2.5\%, for $d=4,   $ it is between 5 and 10\%, and for $d=4,$ it is between 10 and 15\%. The error increases with increasing $\tau$ as the total number of permutations slightly decreases with decreasing $\tau$.}

%%%%% Figure Entropy/Complexity investigation, whole %%%%%
\begin{figure}
\captionsetup[subfigure]{labelformat=empty}
\centering
   \begin{subfigure}[b]{0.99\textwidth}
   \includegraphics[width=0.50\textwidth]{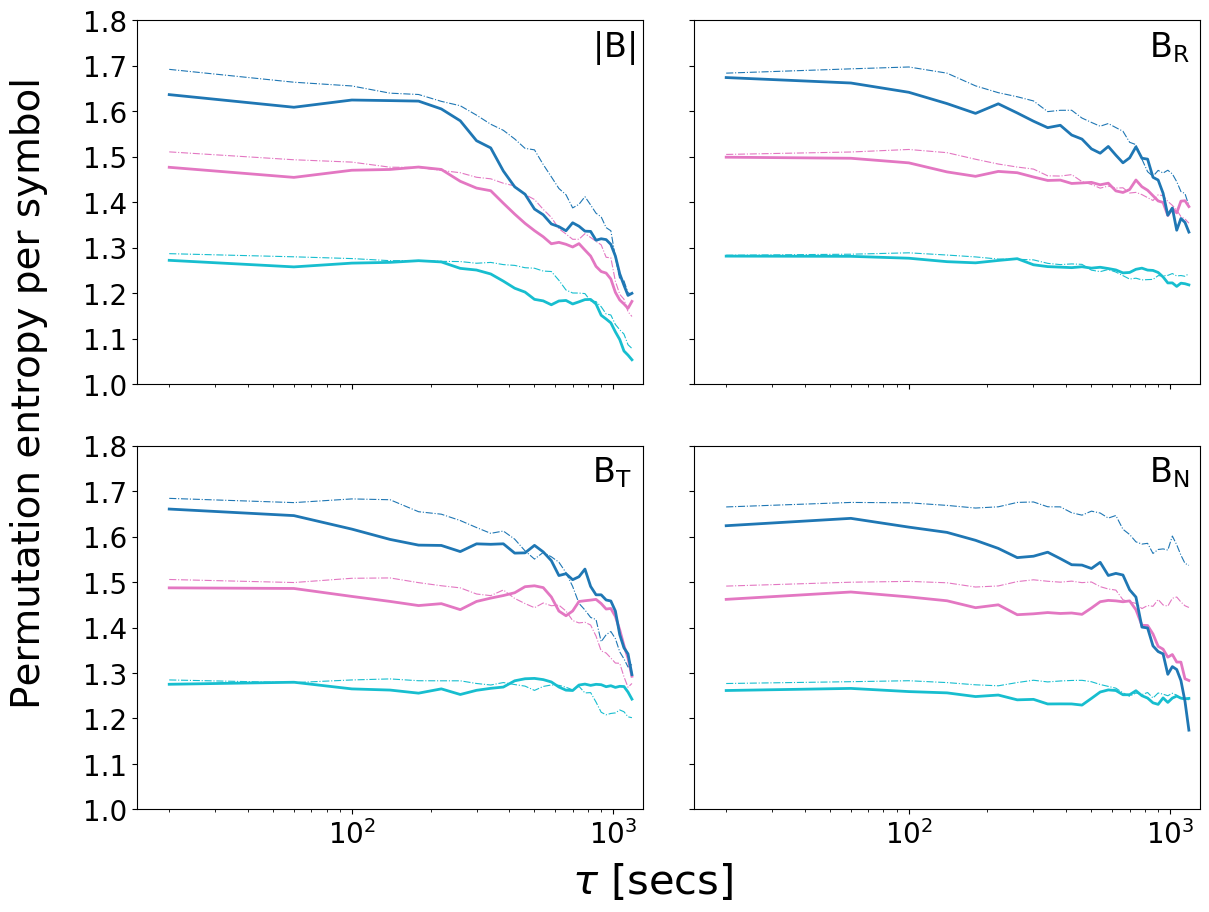}
   \caption{}
\end{subfigure}
   \\
\begin{subfigure}[b]{0.99\textwidth}
   \includegraphics[width=0.50\textwidth]{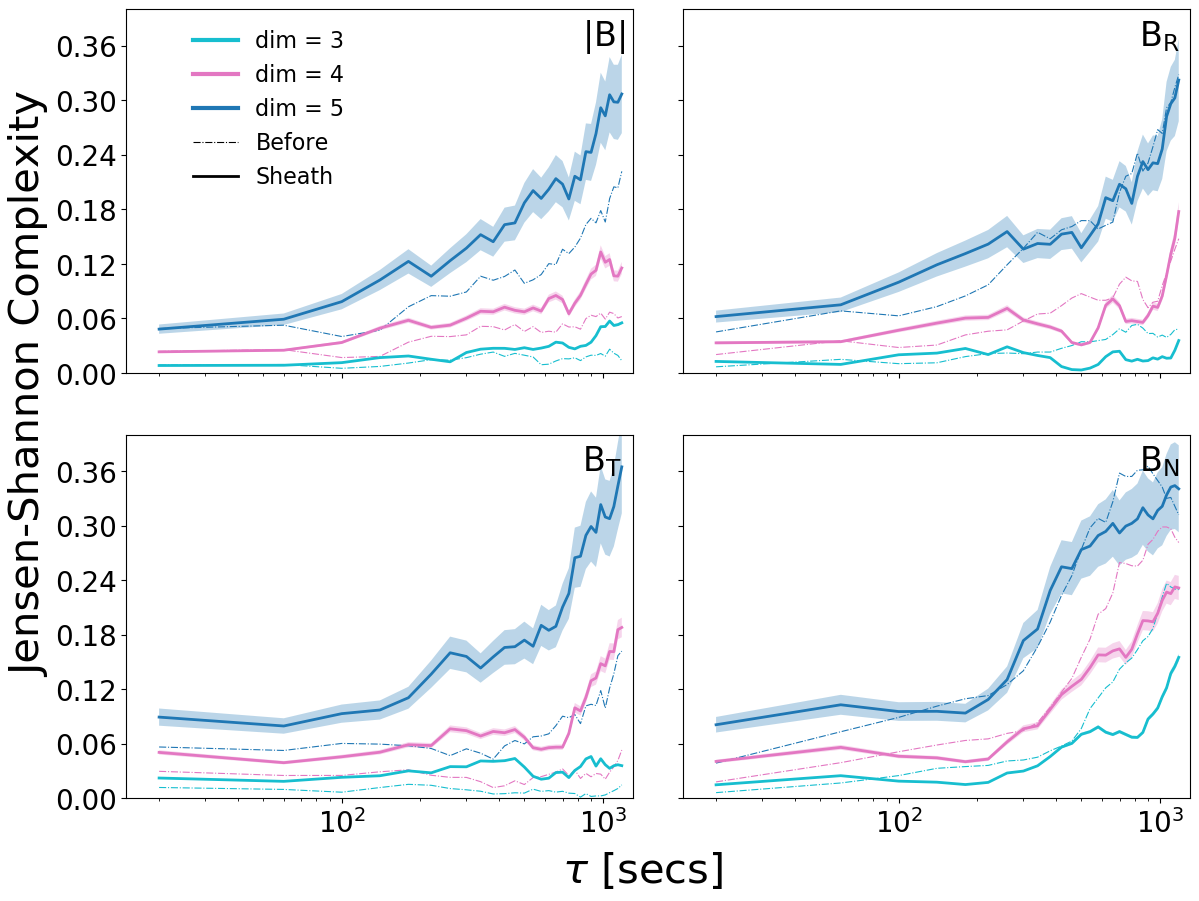}
   \caption{}
\end{subfigure}
\caption{(Top) Permutation entropy per symbol and (Bottom) Jensen-Shannon complexity as a function of timescale $\tau$ for three embedded dimensions ($d=$3, 4, and 5) calculated over 3-hr intervals before the shock and in the sheath (covering almost the whole sheath). The solid thick lines show the results in the sheath, and the dashed lines show the results for the solar wind ahead. In the bottom panels, the uncertainty ranges estimated using the permutation
occupation number approach (see the text) are indicated as shaded areas for the sheath complexity curves. For $d=3, $ the errors are so small that the shaded area is not visible. }
\label{fig:Hn_CH_tau} 
\end{figure}

{Figure~\ref{fig:Hn_CH_tau} shows the permutation entropy per symbol $H_n$ (top) and complexity $C_J^S$ (bottom) as a function timescale $\tau$ for embedded dimensions 3, 4, and 5 in the solar wind preceding the shock (thin dash-dotted lines) and in the sheath (solid thick lines). We note that it is possible that $d=3$ is too low to survey different fluctuation patterns in the solar wind and ICME sheath, but we show it here for completeness. Both regions are 3 hrs in duration, that is, the whole sheath is covered, and $\tau$ ranges from 20 s to 20 min.  The shaded areas for the sheath complexity curves show the uncertainty ranges calculated as previously described. For $d=3,$ the errors are so small that the shaded area is not discernible. 
Some general trends can be distinguished. First, both $H_n$ and $C_J^S$ values consistently increase with the increasing embedded dimension. Secondly, for short timescales, $H_n$ and $C_J^S$ are approximately constant, and for $d=3, $ they both stay relatively flat throughout, except $|B|$ for $H_n$ and $B_N$ for $C_J^S$. For larger embedded dimensions, $H_n$ shows a generally decreasing trend, while the complexity tends to increase with increasing timescale. The uncertainties associated with the complexity curves are generally small and thus indicate that the detected trends probably do not arise from the small sample size. The nearly constant behaviour of $H_n$ and $C_J^S$ as a function of $\tau$ is consistent with stochastic fluctuations \citep[e.g.][]{Osmane2019}. Figure~\ref{fig:Hn_CH_tau} also shows that the permutation entropy per symbol is mostly lower and the complexity is higher in the sheath than in the solar wind ahead, particularly for $d=5$} 

%%%%% Figure Entropy/Complexity investigation %%%%%
\begin{figure}
\captionsetup[subfigure]{labelformat=empty}
\centering
   \begin{subfigure}[b]{0.99\textwidth}
   \includegraphics[width=0.50\textwidth]{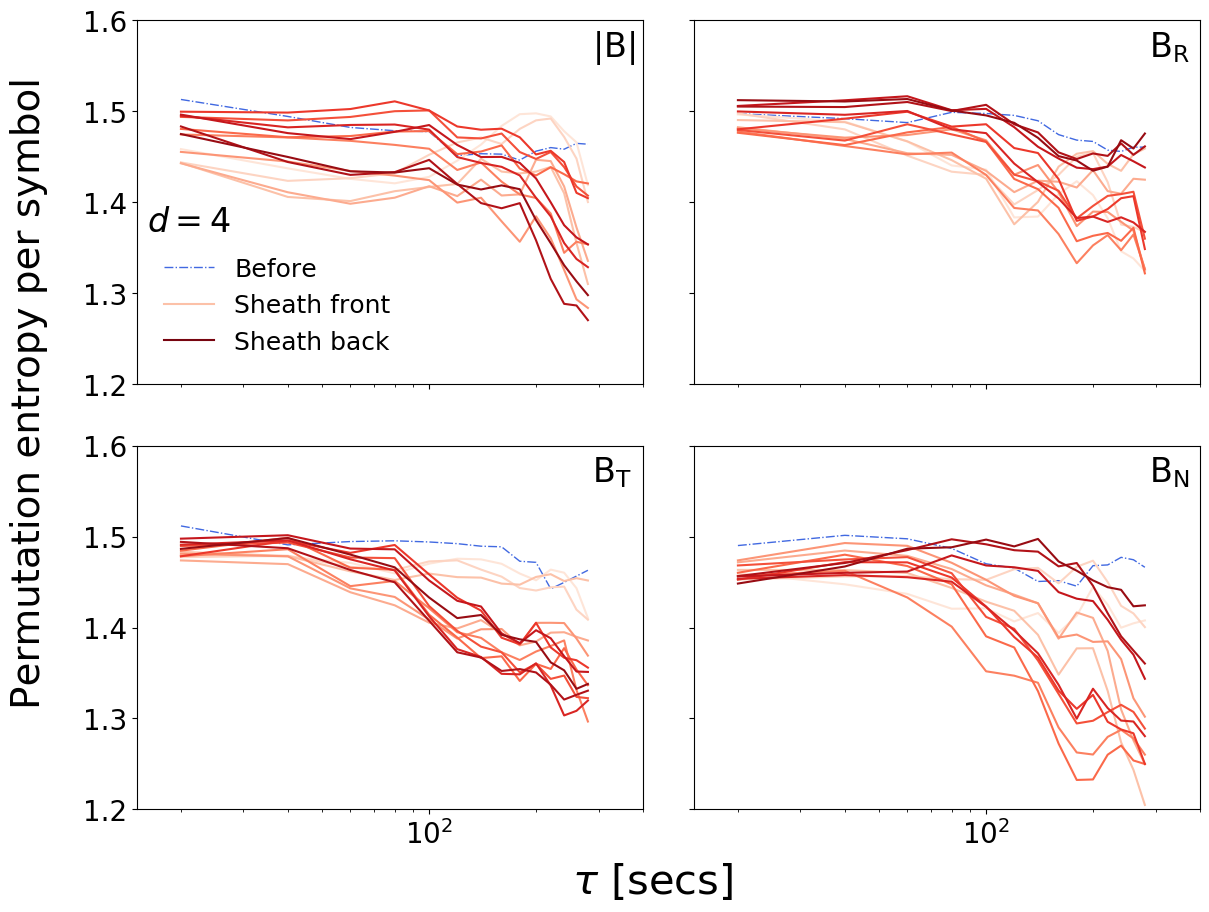}
   \caption{}
\end{subfigure}
   \\
\begin{subfigure}[b]{0.99\textwidth}
   \includegraphics[width=0.50\textwidth]{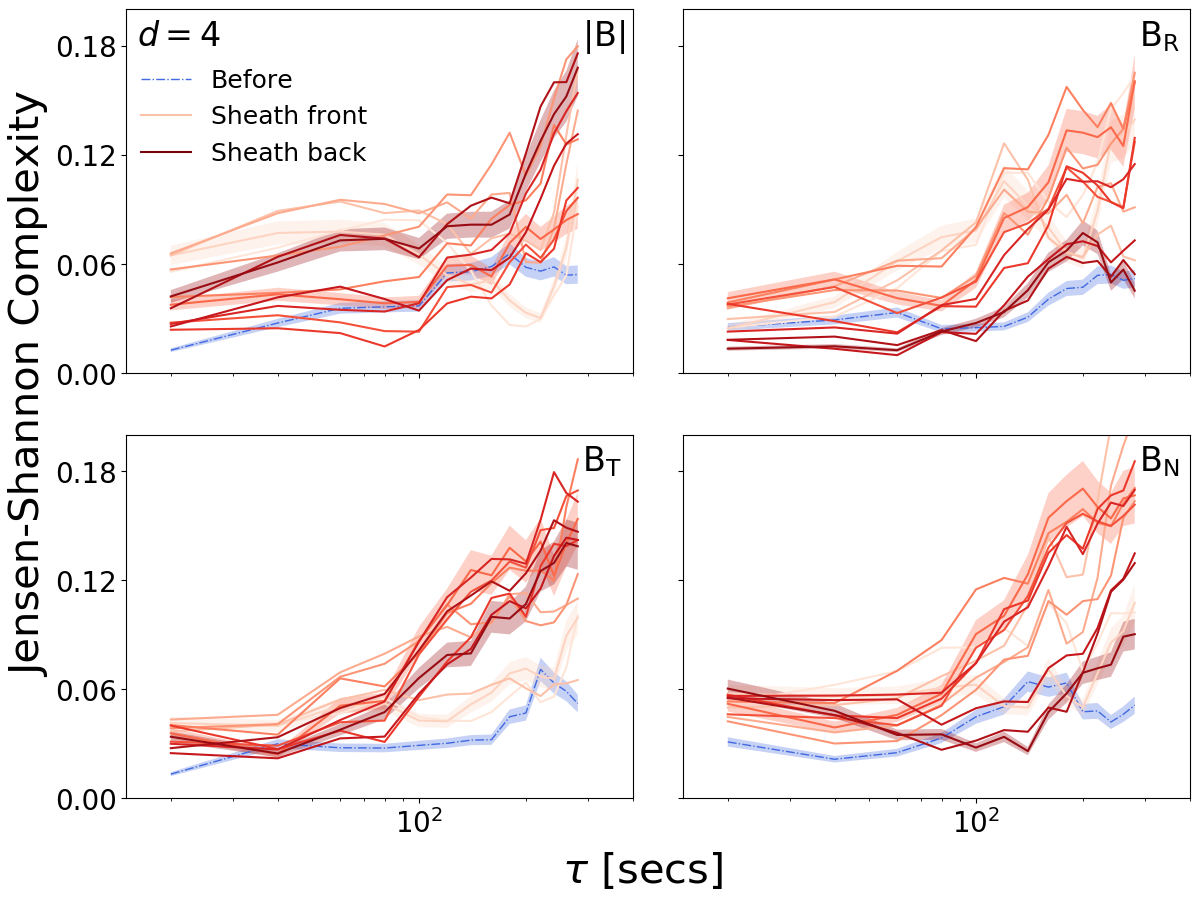}
   \caption{}
\end{subfigure}
\caption{(Top) Permutation entropy per symbol and (Bottom) Jensen-Shannon complexity as a function of timescale $\tau$ for the embedded dimension 4. The solid thick lines show the results in the sheath, and the dashed lines show the results for the solar wind ahead. For the complexity (bottom panel), the uncertainty ranges estimated using the permutation
occupation number approach (see the text) are indicated as shaded areas for the sheath curves.}
   \label{fig:Hn_CH_tau_SLIDE}
\end{figure}

{We further investigated how $H_n$ and $C_J^S$ varied in the sheath by calculating them as a function of timescale for 1-hr intervals, sliding in 15-min steps. The results are shown in Figure~\ref{fig:Hn_CH_tau_SLIDE} (orange and red curves) and are compared to the 1-hr solar wind interval before the shock (blue curves). Again, the shaded areas show the uncertainty range for a few selected curves. We chose to investigate here the embedded dimension $4$, as stated earlier. For larger dimensions, the robustness criteria were not met. Due to the shorter time series compared to the 3-hr time series shown in Figure~\ref{fig:Hn_CH_tau}, the uncertainty ranges are now wider, but some general trends can be distinguished. The plot shows that for most of the sheath subintervals, $H_n$ is smaller and $C_J^S$ is larger for nearly all timescales, in particular for longer timescales, than in the preceding wind. All curves also feature a similar flat profile for short timescales, as in Figure~\ref{fig:Hn_CH_tau}.} 

{The trends how $H_n$ and $C_J^S$ vary within the sheath from the subregions closest to the shock to the ICME leading edge are quite random, in particular for $H_n$. The variations in the $H_n$ and $C_J^S$ values between different sheath subregions are also relatively small for small $\tau$ (but as shown by the uncertainty ranges for $C_J^S$, they are still mostly expected to reflect true variations)  and increase with increasing $\tau$. For longer timescales ($\gtrsim 100$ s), for $B_R$ , the complexity values are highest close to the shock (light orange curves) and decrease deeper within the sheath, where they are almost comparable to the preceding wind curve closest to the ICME leading edge (dark red curves). 
The entropy for $B_R$ in turn is lowest close to the shock and highest close to the ICME leading edge. In contrast, for $B_T$ , the complexity values are first low closest to the shock (light orange curves), then rise in the mid-sheath (dark orange curves), and finally fall again close to the ICME leading edge (dark red curves). Again, the entropy shows the opposite variations within the sheath to complexity.  The entropy for $B_T$ is in turn higher in the front part of the sheath than in the trailing part. The $B_N$ component shows the highest complexity values in the mid-sheath, and similar to $B_R$, it i slowest close to the ICME leading edge, with entropy exhibiting the opposite trend.}

%%%%% figure map %%%%%
\begin{figure}[ht]
\centering
\includegraphics[width=0.99\linewidth]{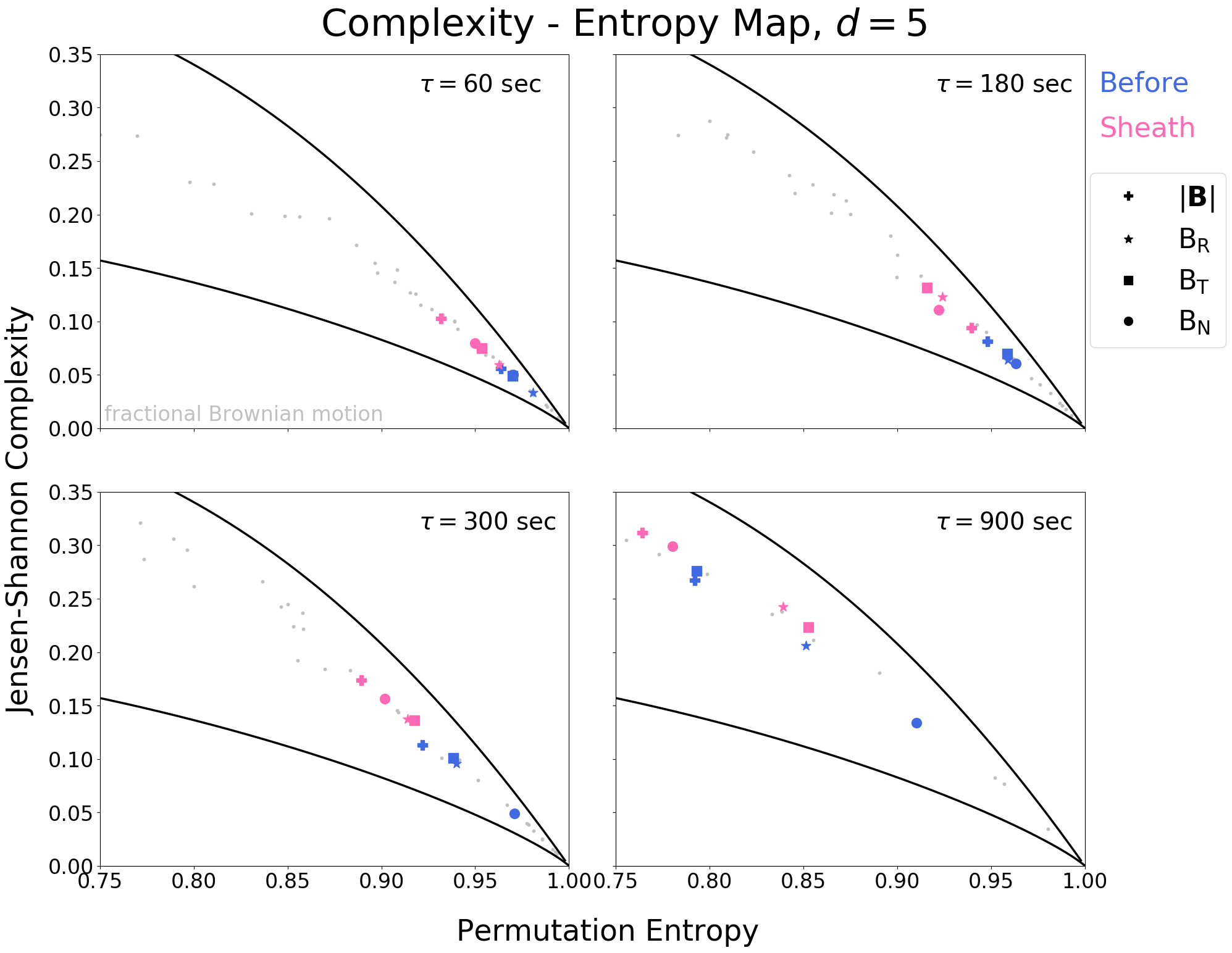}
\caption{Jensen–Shannon permutation entropy - complexity map calculated for 3-hr periods in the upstream and in the sheath. The black curves show the maximum and minimum complexity curves. Grey dots show the stochastic fractional Gaussian motion.}
\label{fig:CH_map}
\end{figure}
%%%%%%%%%%%%%%%%%%%%

{Plotting the calculated CH values against H(P) yields further information on the nature of fluctuations. In this Jensen–Shannon complexity - entropy map, chaotic, stochastic, and periodic fluctuations fall within distinct regions \citep[see e.g. Figure 1 in][]{Weygand2019}. All points in the map are bounded by the maximum and minimum complexity curves, representing the case of a uniform probability distribution and the case where all patterns have equal probability, respectively. Chaotic fluctuations have the highest possible complexity, and they are found close to the top part of the maximum complexity curve. If there is a strong noise component, they move towards the higher-entropy region. Periodic fluctuations in turn are found in the left upper part of the maximum complexity curve, that is, they have medium complexity and low entropy. Stochastic fluctuations are represented by fractional Brownian motion.}

{Figure~\ref{fig:CH_map} shows the lower right part of the Jensen–Shannon complexity - entropy map for $d=5$ for four timescales of the RTN magnetic field components and the magnetic field magnitude. For an example of a complete map and different regions of chaotic, stochastic, and periodic fluctuations, see Figure 1 in \cite{Weygand2019}. Values are calculated for 3-hr periods in the solar wind preceding the shock and in the sheath. The light green curve indicates the minimum complexity curve, the dark green curve shows the maximum complexity curve, and grey points show the fractional Brownian motion. The figure shows that for all timescales, all measurement points overlap with the stochastic fractional Brownian motion. Moreover, they all have relatively low complexity and high entropy. For nearly all points, the sheath points have a slightly higher complexity and lower entropy than in the solar wind ahead. For most cases, magnetic field magnitude and N-component have the highest complexity and lowest entropy.} 

%%%%%%%%%%%%%%%%%%%%%%%%%%%%%%%%%%%%%%%%
%%% DISCUSSION
%%%%%%%%%%%%%%%%%%%%%%%%%%%%%%%%%%%%%%%%

\section{Discussion} \label{sec:discussion}

{The investigated ICME sheath showed enhanced levels of magnetic fluctuations (both $\delta B$ and $\delta B /B$) and a significantly higher occurrence frequency of coherent structures (PVI $> 3$) than the preceding wind. These trends are consistent with previous studies \citep{Moissard2019,Good2020a,Kilpua2020,Kilpua2021} that have related them to sheaths amplifying upstream fluctuations and locally generating new ones. Our study revealed that enhanced levels of turbulence were maintained throughout the sheath: All distributions of $\delta B$, $\delta B /B,$ and PVI calculated using a sliding 1-hr window from the shock to the ICME leading edge showed extended tails at high values that were absent from the upstream wind.}

{The findings in this work are interesting as the ICME driving the sheath was detected at an earlier evolutionary stage than in most previous studies ($\sim 0.5$ au compared to 1 au) and because the driving CME was slow and slowly expanding, and was associated with a pair of relatively weak shocks \citep{Lario2020}. Enhanced turbulence  in a younger sheath was also reported by \cite{Good2020a}, who analysed a sheath ahead of a slow CME observed by MESSENGER at $0.47$ au. That sheath was also encountered by almost radially aligned STEREO-B at $\sim 1$ au where a far smaller difference in fluctuation properties between the sheath and the preceding wind was detected.}

{The results suggest that for the event we studied and for the fluctuation timescales we analysed, the relatively weak ICME leading shock(s) had a minor effect on the preceding wind, and most processing and generation of fluctuations occurred within the sheath.  While several studies \citep[e.g.][]{Zhao2019b,Zhao2019a,Zank2021, Borovsky2020,Zhao2020,zhao2021} have reported fluctuation amplitudes of velocity, density, and magnetic field  to  amplify downstream of the shock, at heliospheric distances ranging from ~1 au to 84 au, we observed here the most significant fluctuation enhancements and the fattest tails of $\delta B$, $\delta B /B,$ and PVI deeper in the sheath. The  sheath had several clear and sharp field changes, a few of which presumably were HCS crossings  swept and compressed into the sheath, that were associated with particularly enhanced fluctuations and large PVI values. The finding that the magnetic compressibility of fluctuations was similar to the upstream wind suggests that new compressible fluctuations were not significantly generated in this sheath at the investigated timescales. This is consistent with the general shock studies referred to above, despite the compression at the shock.}

{We also noted that a larger increase in $\delta B / B$ with increasing timescale in the sheath when compared to preceding wind suggest steeper MHD range spectral slopes in the sheath. This is consistent with \cite{Kilpua2021}, who reported that the majority of slow sheaths have spectral slopes significantly steeper than Kolmogorov slopes, and that slopes are generally steeper in the sheath than in the upstream.  This also agrees with our finding of a higher frequency of high PVI values in the sheath as the presence of coherent structures steepens the spectral slopes \citep[e.g.][]{Li2012,Borovsky2020}.}

{In the sheath we studied, generally negative residual energy $\sigma_r$ indicated an excess of energy in magnetic field fluctuations, and generally positive cross helicity $\sigma_c$ indicated that more power was in waves propagating away from the Sun (given that the mean field was in the toward sector). These trends are consistent with findings by \cite{Good2020b} and \cite{Zhao2020b}, who analysed an ICME and surrounding wind detected by the PSP on November 12, 2018, at 0.25 au. This ICME did not drive a shock, but had an extended perturbed upstream region, that is, a shockless sheath. Sheath fluctuations were found generally to show relatively low Alfv\'enicity and to be dominated by  anti-parallel propagating waves corresponding to propagation away from the ICME, that is, away from the Sun, consistent with general characteristics in the solar wind beyond the Alfv\'en critical point. However, \cite{Good2020b} reported  highly Alfv\'enic fluctuations in the outer layers of the driving ICME. In our study, we also detected $\sigma_r$  closest to $0$ in the vicinity of the ICME leading edge, as well as close to the shock.}

\begin{figure}[ht]
\centering
\includegraphics[width=0.99\linewidth]{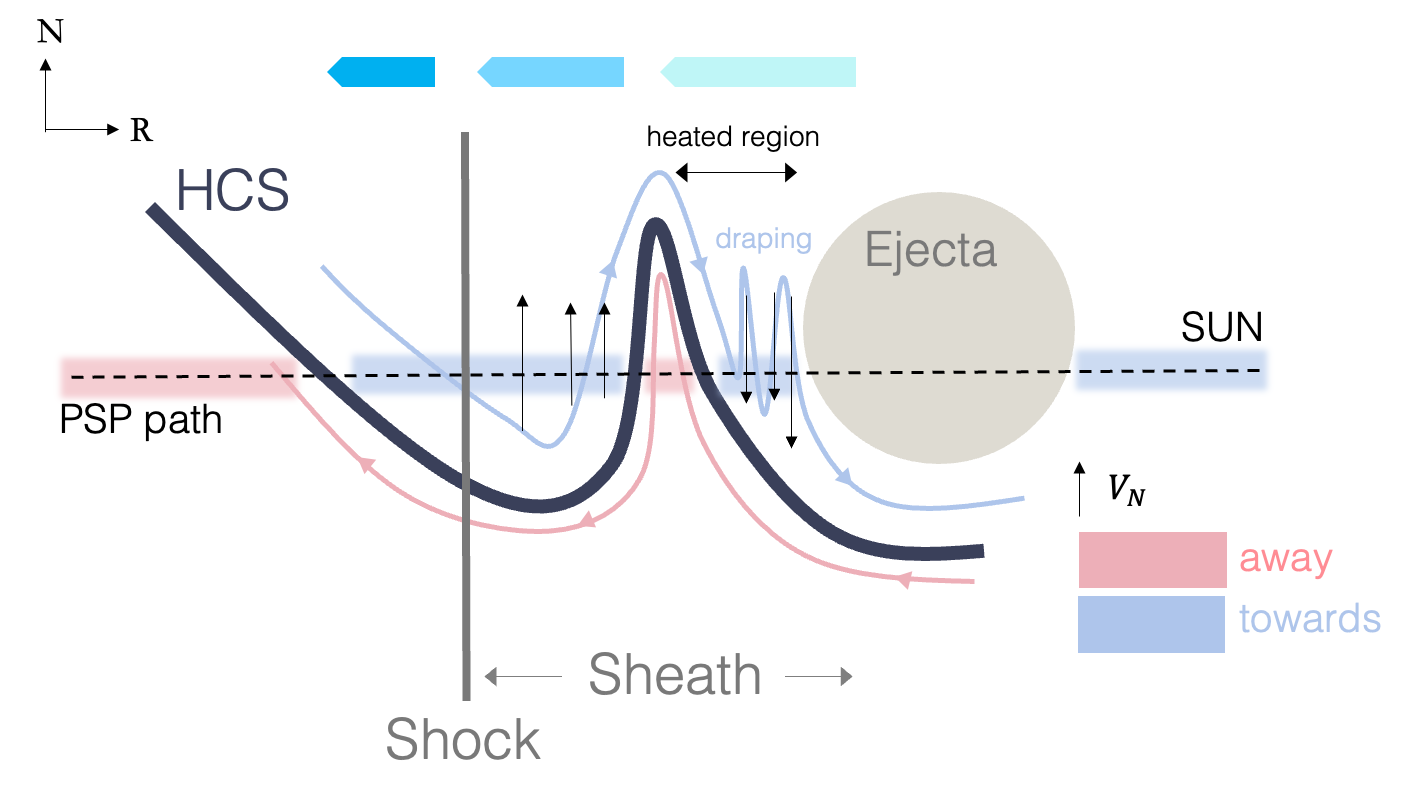}
\caption{Sketch of the sheath, ejecta, and the warped HCS (not to scale). The field lines are projections in the R-N plane. Blue and red areas indicate directions away and towards the magnetic sectors. Blue arrows at the top show the relative solar wind speed in the upstream and in the sheath, before and after the warp. The longer and lighter coloured the arrow, the faster the speed.}
\label{fig:sketch}
\end{figure}

{The investigated sheath featured a two-step speed profile instead of the linear profiles observed typically at 1 au \citep[e.g.][]{Kilpua2019,Salman2021}.  The speed gradient occurred during the HCS crossings and coincided with a velocity shear zone (in $V_N$ and  $V_T$). Figure~\ref{fig:sketch} shows a sketch of the driving ICME (ejecta), sheath, and HCS. The HCS likely formed a warp that was swept up and compressed into the sheath. The ICME is marked to be crossed slightly below its axis, as suggested by its northward propagation from the ecliptic by \cite{Lario2020}. Velocity shear occurs across this warp, as indicated for $V_N$ by thin black arrows in Figure~\ref{fig:sketch}. The HCS crossings marked significant increases in temperature and specific entropy. We suggest that the ICME accelerated and heated the solar wind between its leading edge and the HCS warp, and that this faster flow interacted with the slower wind ahead, creating a structure resembling a stream interface \citep[SI; e.g.][]{Borovsky2010}. The faster flow after the HCS warp was also related to most enhanced fluctuations properties ($\delta B$, $\delta B/B,$ and PVI), likely related to the interaction and field draping around the ICME. $\sigma_r$ and $\sigma_c$ deviated also from the general trends in the sheath at this time.  In the shear zone, the strongly negative $\sigma_r$ (and the low Alfv\'en ratio) suggests a particularly strong excess of energy in magnetic fluctuations and indicates that the small-scale structures are formed in the vicinity of HCS \citep{zhao2021}.  The faster flow after the HCS warp  showed localised patches of strongly positive $\sigma_r$, that is, excess of energy in velocity fluctuations. Varying $\sigma_r$ have been reported in stream interaction regions in previous studies \citep[e.g.][]{Shi2020}.  We note that \cite{Kilpua2021b} reported that a localised HCS warp in the sheath of an April 2020 ICME at L1 was related to an enhancement of energetic ions, but for the studied event, no energetic ion enhancement was observed in the sheath \citep{Lario2020}. Switchbacks are also frequently detected structures in the PSP data at its closer passages to the Sun, which can cause abrupt and large changes in the magnetic field direction \citep[e.g.][]{Bale2019}. We consider it unlikely, however, that the structure observed here in the middle of the sheath is a switchback. $V_R$ does not show a jet-like profile, but rather, it first increases and then stays elevated after the later boundary crossing, the largest field variation is in $B_N$, and the heat flux  changes direction in the structure.}

%%%%% figure map %%%%%
%%%%%%%%%%%%%%%%%%%%

{Only a few studies have investigated the nature of  solar wind fluctuations using the complexity-entropy method  \citep[][]{Weck2015,Weygand2019,Good2020a}. All these previous studies have found solar wind fluctuations to be stochastic given that they occupy the lower-right region of the complexity-entropy plane and overlap with fractional Brownian motion. We also observed the same behaviour. As proposed by \cite{Weck2015}, this reflects that the solar wind can in general be described as fully developed turbulence. We found that both in the sheath and in the preceding winds, the permutation entropy and complexity stayed approximately constant at short timescales, while entropy clearly decreased and complexity increased with increasing timescale. However, the analysis showed that sheath fluctuations had a slightly higher complexity and lower entropy than the preceding slow wind. These findings agree with  \cite{Good2020a},
who suggested that a higher complexity of sheath fluctuations is consistent with the current understanding how sheaths form, that is, as a variable mix of coherent, ordered structures and random, disordered fluctuations. The results from the present work further demonstrate that these features hold for different embedded dimensions ($d=3,4,$ and 5) and for different subregions in the sheath (for $d=4$). The behaviour of entropy and complexity as a function of timescale is opposite to the geomagnetic AE and AL indices studied by \cite{Osmane2019}. This could be due to intermittent coherent structures at longer timescales having a greater importance for solar wind, while geomagnetic disturbances at longer timescales are likely more random than at shorter timescales. }

{Permutation entropy and complexity curves as a function of timescale did show some variation throughout the sheath. These variations could have been related to the presence of coherent intermittent structures. In particular, $B_N$ showed highest complexity and lowest entropy in the middle of the sheath, reflecting the presence of the HCS warp and associated fluctuations. The analysis of a larger sample of sheaths is required to determine if there are any common trends, however. We note that the duration of this sheath, which was observed for only $\sim 3$ hrs, limited the range of fluctuation timescales that could be studied, and it limited the maximum embedded dimension to $d=5$. We thus cannot rule out that higher dimensions would show chaotic or periodic fluctuations in the CME-driven sheath.}

%%%%%%%%%%%%%%%%%%%%%%%%%%%%%%%%%%%%%%%%
%%% CONCLUSION
%%%%%%%%%%%%%%%%%%%%%%%%%%%%%%%%%%%%%%%%

\section{Conclusion} \label{sec:conclusion}

{We have studied the structure and fluctuation properties of a sheath region observed by the PSP on March 15, 2019, at $\sim 0.55$ au, ahead of a slowly propagating and slowly expanding ICME associated with a streamer blow-out CME preceded by a pair of relatively weak shocks. The sheath included a warped HCS that divided it into two different flows.  The cross-helicity and normalised residual energy analysis suggest that sheaths, like the solar wind, are generally magnetically dominated structures where fluctuations propagate away from the Sun, but with significant local variations. Our results (and those by \cite{Good2020a}) suggest that closer to the Sun, sheaths could already frequently include coherent intermittent structures such as current sheets, sharp field discontinuities, large-angle field rotations, and magnetic holes. In slow CME sheaths, enhanced fluctuations are thus likely rather associated with the presence and processing of such structures from the preceding wind than with the active generation of new fluctuations. Further observations closer to the Sun by the PSP, Solar Orbiter, and BepiColombo will allow examination of typical features in younger sheaths.}

%%%%%%%%%%%%%%%%%%%%%%%%%%%%%%%%%%%%%%%%
%%% ACKNOWLEDGEMENTS
%%%%%%%%%%%%%%%%%%%%%%%%%%%%%%%%%%%%%%%%

\begin{acknowledgements}
{The results presented here have been achieved under the framework of the Finnish Centre of Excellence in Research of Sustainable Space (FORESAIL; Academy of Finland grant numbers 312390 and 336809), which we gratefully acknowledge. EK acknowledges the ERC under the European Union's Horizon 2020 Research and Innovation Programme Project 724391 (SolMAG), and EK and SG acknowledge Academy of Finland Project 310445 (SMASH). This study has also been partially funded through the European Union’s Horizon 2020 research and innovation programme under grant agreement No 101004159 (SERPENTINE). Data analysis was performed with the AMDA science analysis system provided by the Centre de Données de la Physique des Plasmas (CDPP) supported by CNRS, CNES, Observatoire de Paris and Université Paul Sabatier, Toulouse. The FIELDS experiment on the Parker Solar Probe spacecraft was designed and developed under NASA contract NNN06AA01C. We acknowledge the NASA Parker Solar Probe Mission and the SWEAP team led by J. Kasper for use of data.} 
\end{acknowledgements}

% for the bibliography, at the end
\bibliographystyle{aa} % style aa.bst
\bibliography{Bib_SH.bib} % your references file.bib

\begin{appendix} 

\section{Reconnection exhausts} 
\label{app:a}

%%%%% Appendix figure 1 %%%%%
\begin{figure}[ht]
\centering
\includegraphics[width=0.99\linewidth]{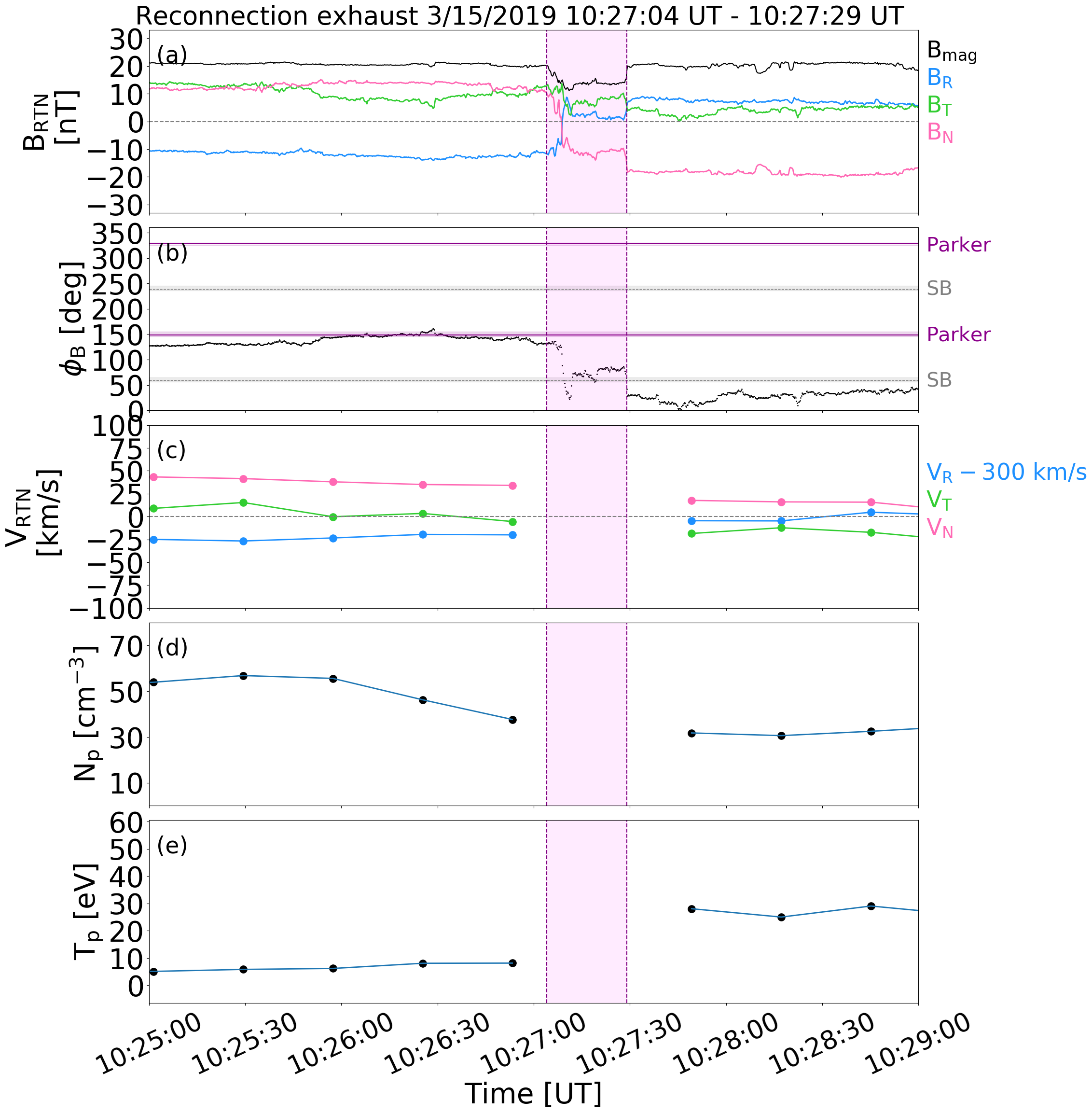}
\caption{Solar wind magnetic field and plasma data as measured by the PSP during the first reconnection exhaust within the sheath (the first magenta line in Figure~\ref{fig:overview}). From top to bottom: a) Magnetic field magnitude (black) and three magnetic field components in RTN (blue: $B_R$, green: $B_T$, pink: $B_N$), b) azimuth RTN angle of the magnetic field, c) three solar wind velocity  components in RTN (blue: $V_R - 300$ km/s, green: $V_T$, pink: $V_N$, d) solar density and e) solar wind temperature.}
\label{fig:Exhaust1}
\end{figure}

%%%%% Appendix figure 2 %%%%%
\begin{figure}[ht]
\centering
\includegraphics[width=0.99\linewidth]{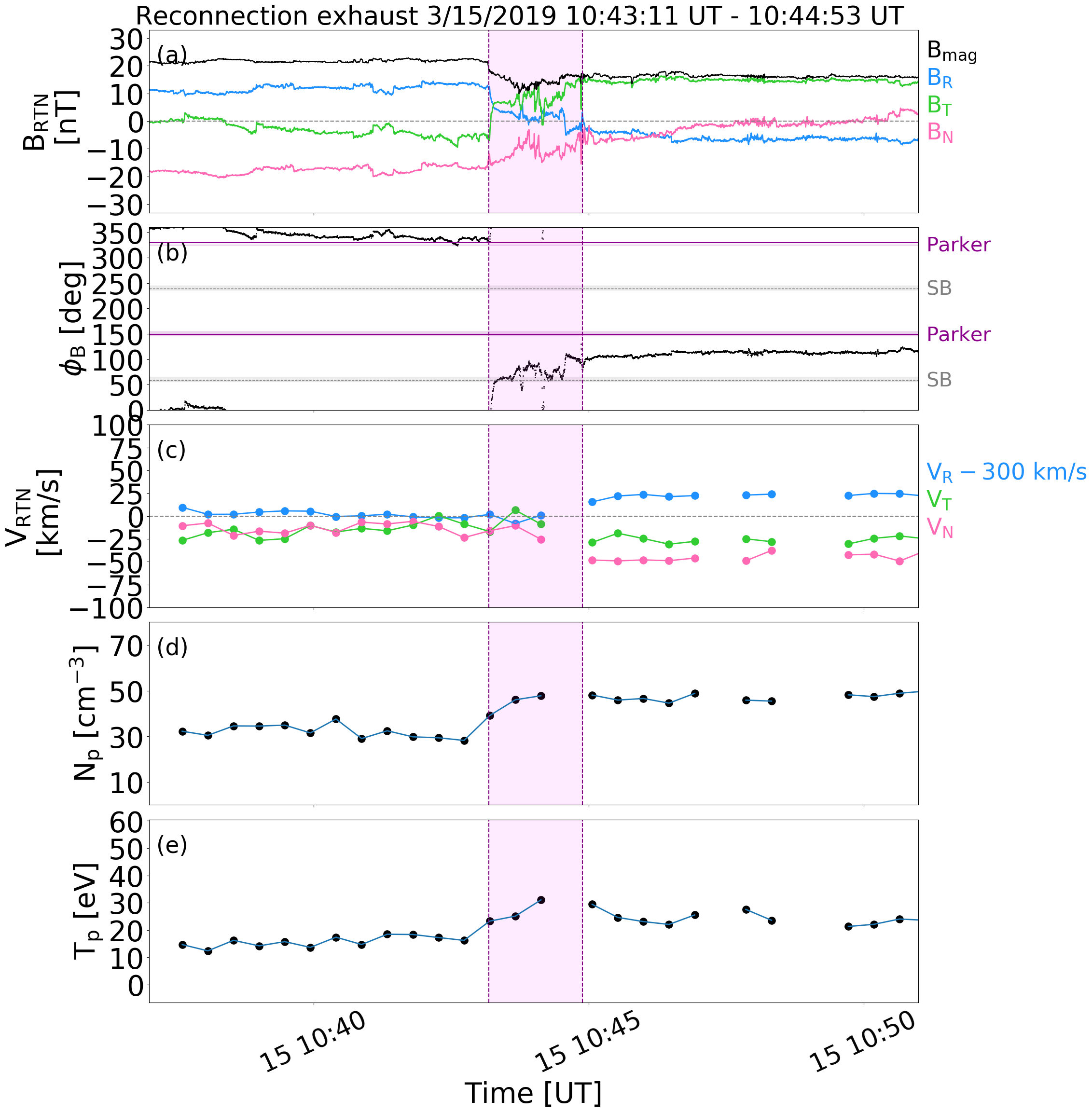}
\caption{Solar wind magnetic field and plasma data as measured by the PSP during the second likely reconnection exhaust within the sheath (the second magenta line in Figure~\ref{fig:overview}), in the same format as Figure~\ref{fig:Exhaust1}.}
\label{fig:Exhaust2}
\end{figure}

\end{appendix} 
%%%%%%%%%%%%%%%%%%%%

\end{document}